\documentclass{acm_proc_article-sp}
\usepackage{graphicx}
\usepackage{array}
\usepackage{subfigure}
\usepackage{amsmath}
\usepackage{url}
\usepackage{color}
\usepackage{multirow}
\usepackage{hhline}
\makeatletter
\def\ps@headings{%
\def\@oddhead{\mbox{}\scriptsize\rightmark \hfil \thepage}%
\def\@evenhead{\scriptsize\thepage \hfil \leftmark\mbox{}}%
\def\@oddfoot{}%
\def\@evenfoot{}}
\makeatother
\def\_{\rule{.3em}{.15ex}}      % Get underscore by typing \_.

 \normalmarginpar
\newcommand {\mymarginpar}[1]{\marginpar{#1}}
\renewcommand {\marginpar}[1]{}

\def\_{\rule{.3em}{.15ex}}      % Get underscore by typing \_.

\newcommand{\ls}[1]
   {\dimen0=\fontdimen6\the\font
    \lineskip=#1\dimen0
    \advance\lineskip.5\fontdimen5\the\font
    \advance\lineskip-\dimen0
    \lineskiplimit=.9\lineskip
    \baselineskip=\lineskip
    \advance\baselineskip\dimen0
    \normallineskip\lineskip
    \normallineskiplimit\lineskiplimit
    \normalbaselineskip\baselineskip
    \ignorespaces
   }
\newcommand {\bearn}{\begin{eqnarray*}}
\newcommand {\eearn}{\end{eqnarray*}}
\newcommand {\barr}{\begin{array}}
\newcommand {\earr}{\end{array}}

\newcommand {\N}{{\cal N}}

\renewcommand {\S}{{\cal S}}

\newtheorem{definition}{Definition}
\newtheorem{property}[definition]{Property}
\newtheorem{proposition}[definition]{Proposition}
\newtheorem{lemma}[definition]{Lemma}
\newtheorem{theorem}[definition]{Theorem}
\newtheorem{corollary}[definition]{Corollary}
\newtheorem{example}[definition]{Example}
\newtheorem{remark}[definition]{Remark}
\newcommand {\benum} {\begin{enumerate}}
\newcommand {\eenum} {\end{enumerate}}

\newcommand {\bdesc} {\begin{description}}
\newcommand {\edesc} {\end{description}}

\newcommand {\bfig}[2] {\begin{figure}
  \centering
  \includegraphics[width=#2]{#1}}
\newcommand {\brotatefig}[2] {\begin{figure}[htbp]
                        \centerline {
                         \epsfig{figure={#1},clip=,angle=-90,width={#2}}}}
\newcommand {\bfigfirst}[2] {\begin{figure}[h]
                        \centerline {
                        \setlength{\epsfxsize}{#2}
                        \epsffile{#1}}}
\newcommand {\efig}[2]{ \caption{#2}
                        \label{fig:#1}
                        \end{figure}
                        \mymarginpar{fig:#1}}
\newcommand {\erotatefig}[2]{ \caption{#2}
                        \label{fig:#1}
                        \end{figure}
                        \mymarginpar{fig:#1}}

\newcommand {\btab}[1]{
                       \begin{table}
                       \centering
                       \begin{tabular}{#1}}
\newcommand {\etab}[3] {
                       \end{tabular}
                       \caption[#3]{#2}
                       \label{tab:#1}
                       \end{table}
                       \mymarginpar{tab:#1}
                       \vspace{.1in}}

\newcommand {\btabular}[1]{\begin{center}
                       \begin{tabular}{#1}}
\newcommand {\etabular}{\end{tabular}
                       \end{center}}

\newcommand {\bdefin}[1]{\begin{definition}
                      \mymarginpar{def:#1}
                      \label{def:#1} }
\newcommand {\edefin}       {\end{definition}}

\newcommand {\bpro}[1]{\begin{property}
                      \mymarginpar{pro:#1}
                      \label{pro:#1} }
\newcommand {\epro}   {\end{property}}

\newcommand {\bprop}[1]{\begin{proposition}
                      \mymarginpar{prop:#1}
                      \label{prop:#1} }
\newcommand {\eprop}       {\end{proposition}}

\newcommand {\blem}[1]{\begin{lemma}
                      \mymarginpar{lem:#1}
                      \label{lem:#1} }
\newcommand {\elem}   {\end{lemma}}

\newcommand {\bthe}[1]{\begin{theorem}
                      \mymarginpar{the:#1}
                      \label{the:#1} }
\newcommand {\ethe}   {\end{theorem}}

\newcommand {\bcor}[1]{\begin{corollary}
                      \mymarginpar{cor:#1}
                      \label{cor:#1} }
\newcommand {\ecor}   {\end{corollary}}

\newcommand {\bax}[1]{\begin{axiom}
                      \mymarginpar{ax:#1}
                      \label{ax:#1} }
\newcommand {\eax}       {\vspace{-.1in} \end{axiom}}

\newcommand {\bex}[2]{\vspace{.1in}
                      \begin{example}
                      \mymarginpar{ex:#1}
                       {\bf #2}
                      \label{ex:#1} }
\newcommand {\eex}       {\end{example} \vspace{.3cm} }

\newcommand {\brem}[1]{\begin{remark}
                      \mymarginpar{rem:#1}
                      \label{rem:#1} \em }
\newcommand {\erem}   {\end{remark}}

\newcommand {\beq}[1]{\mymarginpar{eq:#1}
                      \begin{equation}
                      \label{eq:#1} }

\newcommand {\beqno}[1]{\mymarginpar{eq:#1}
                      \begin{eqnarray}
                      \nonumber}

\newcommand {\eeq}       {\end{equation}}
\newcommand {\eeqno}       { && \end{eqnarray}}

\newcommand {\bear}[1]{\mymarginpar{eq:#1}
                       \begin{eqnarray}
                       \label{eq:#1} }

\newcommand {\bearno}[1]{\mymarginpar{eq:#1}
                       \begin{eqnarray}
                       \nonumber}

\newcommand {\eear}{\end{eqnarray}}
\newcommand {\eearno}{\end{eqnarray}}
\newcommand {\bsel}{\left \{ \begin{array}{cl}}
\newcommand {\esel}{\end{array} \right.}

\newcommand {\bmat}[1]{\left [ \begin{array}{#1}}
\newcommand {\emat}{\end{array} \right ]}
\def\R{I\kern-0.30em R}
\def\N{I\kern-0.30em N}
\def\P{I\kern-0.30em P}

\begin{document}
\title{Temporal Matrix Factorization for Tracking Concept Drift in Individual User Preferences}

\iffalse
\author{\IEEEauthorblockN{Yung-Yin Lo}
\IEEEauthorblockA{Graduated Institute of Electrical\\ Engineering\\
National Taiwan University\\
Taipei, Taiwan, R.O.C.\\
Email: r02921080@ntu.edu.tw}
%ying2716@gmail.com}
\and
\IEEEauthorblockN{Wanjiun Liao}
\IEEEauthorblockA{Department of Electrical \\ Engineering\\
National Taiwan University\\
Taipei, Taiwan, R.O.C.\\
Email: wjliao@ntu.edu.tw}
\and
\IEEEauthorblockN{Cheng-Shang Chang}
\IEEEauthorblockA{Institute of Communications\\ Engineering\\
National Tsing Hua University\\
Hsinchu 300, Taiwan, R.O.C.\\
Email: cschang@ee.nthu.edu.tw}}
\fi

\numberofauthors{3}

\author{
\alignauthor Yung-Yin Lo\\
       \affaddr{Graduated Institute of Electrical}\\
       \affaddr{Engineering}\\
      \affaddr{National Taiwan University}\\
      \affaddr{Taipei, Taiwan, R.O.C.}\\
       \email{r02921080@ntu.edu.tw}
\alignauthor Wanjiun Liao\\
       \affaddr{Graduated Institute of Electrical}\\
       \affaddr{Engineering}\\
       \affaddr{National Taiwan University}\\
       \affaddr{Taipei, Taiwan, R.O.C.}\\
       \email{wjliao@ntu.edu.tw}
\alignauthor Cheng-Shang Chang\\
       \affaddr{Institute of Communications}\\
       \affaddr{Engineering}\\
       \affaddr{National Tsing Hua University}\\
       \affaddr{Hsinchu 300, Taiwan, R.O.C.}\\
       \email{cschang@ee.nthu.edu.tw}
}

\maketitle

\begin{abstract}
%\boldmath
The matrix factorization (MF) technique has been widely adopted for solving the rating prediction problem
in recommender systems. The MF technique utilizes the latent factor model to obtain {\em static} user preferences (user latent vectors) and item characteristics (item latent vectors) based on historical rating data. However, in the real world user preferences are not static but full of dynamics. Though there are several previous works  that addressed this time varying issue of user preferences, it seems (to the best of our knowledge) that none of them  is specifically designed for tracking concept drift in {\em individual user preferences}.
             Motivated by this, we develop a Temporal Matrix Factorization approach (TMF) for tracking concept drift in each individual user latent vector. There are two key innovative steps in our approach: (i) we develop a modified stochastic gradient descent method to learn an individual user latent vector at each time step, and (ii) by the Lasso regression we learn a linear model for the transition of the individual user latent vectors. We test our method  on a synthetic dataset and several real datasets. In comparison with the original MF, our experimental results show that our temporal method is able to achieve lower root mean square errors (RMSE) for both the synthetic and real datasets. One interesting finding is that the performance gain in RMSE is mostly from those users
              who indeed have concept drift in their user latent vectors at the time of prediction. In particular, for the synthetic dataset and the Ciao dataset, there are quite a few users with that property and the performance gains for these two datasets are roughly 20\% and 5\%, respectively.

%  by 20\% on the synthetic dataset and 5\% on Ciao real dataset which is a significant result toward this problem.
\end{abstract}

\category{H3.3}{Information Storage and Retrieval}{Information
Search and Retrieval--Information filtering}
\terms{Algorithms; Experimentation}

%\IEEEpeerreviewmaketitle

%\section{Introduction}
%\label{c:introduction}

\section{Introduction}
\label{c:intro}

With the accelerated growth of the Internet and a wide range of web services such as electronic commerce and online video streaming, people are easily overwhelmed by massive amounts of information and therefore recommender systems are indispensable tools to alleviate the information overload problem.
%Recommender systems provide personalized recommendations that enable users to quickly locate desirable items. On %the other hand, they are also an important factor in determining the success of a business. Amazon said that 35\% %of product sales resulted from its recommender system \cite{marshall2006aggregate} and Netflix also reported that %75\% of what its users watched came from recommendations \cite{amatriain2012netflix}. As recommender systems can %benefit both users and providers of online service, they play a much more significant role in this era of big data.
At the heart of each recommender system, there is an algorithm that handles the rating prediction task and the accuracy of the rating prediction algorithm is the foundation of the system. The most successful and widely used approach to implement such an algorithm is collaborative filtering with matrix factorization (MF). Such an approach has the advantage of high accuracy, robustness and scalability, and  it is thus more favorable than the other approaches,  such as the neighborhood-based approach and the graph-based approach \cite{su2009survey, shi2014collaborative}. The MF approach proved its success in the Netflix Prize competition \cite{amatriain2013mining}
%since it achieved the best performance without any hybrid method and even
as the winning submission of this competition was heavily relied on it to predict unobserved ratings. The MF approach decomposes a user-item rating matrix into two low-rank matrices which directly profile {\em users} and {\em items} to the latent factor space respectively and these representative latent factors form the main basis for further prediction in the future.

Although MF is the state-of-the-art approach that can  successfully process the relational rating data, its capability of capturing the temporal dynamics of users' preferences is quite limited.  As we are facing the fast-moving business environment, the real world is not static but full of dynamics. There are a great variety of sources that can cause the changes of users' behavior, including shifting trends in the community, the arrival of new products, the changes in users' social networks, and so on. Recent research in \cite{mcauley2013amateurs} considered the aspect of personal development and pointed out that user's expertise may change from amateurs to connoisseurs as they become more experienced. To satisfy users' current taste and need, a key building block for recommender systems is to accurately model such user preferences as they evolve over time.

The need to model the temporal dynamics of user preferences raises some fundamental challenges. First of all, the amount of available data is significantly reduced in a specific time step and the sparsity problem of recommender systems is more severe in this situation. In addition, how can we generally incorporate the temporal dimension and further capture the evolution of preferences at the individual level for every time step? Finally, what is the principled method to model this kind of transition for every user in order to make more accurate predictions in the future? Toward this end, we propose a general and principled temporal dynamic model for tracking concept drift in each individual user latent vector. Such an approach can further effectively and efficiently achieve a lower RMSE than that of MF.

The main contributions of this paper include:
\begin{itemize}
\item We propose a Temporal Matrix Factorization approach (TMF) for tracking concept drift in each individual user latent vector. Such a method
not only breaks the limit of using static decompositions in the original MF approach, but also provides a tool for recommender systems to  better serve ``valuable'' customers in the future.

\item  We develop a modified stochastic gradient descent \break method to learn an individual user latent vector at each time step by using both the overall rating logs and the rating logs within the specific time step.

\item  By using the Lasso regression for  the user latent vector at every time step, we learn a linear system model that can be used for modelling the transition pattern at the individual level.

 \item We conduct comprehensive experiments on a synthetic dataset and four real datasets,  Ciao, Epinions, Flixster and MovieLens. In comparison with the original MF, our experimental results show that our TMF approach is able to achieve lower root mean square errors (RMSE) for both the synthetic and real datasets.
               In particular, there is roughly a 17-26\% improvement on the synthetic dataset and a 5\% improvement on the Ciao dataset. Such an improvement is quite significant.
%                in view of the three-year Netflix Prize competition to improve RMSE by only 10\%.

\item  Our experiments also reveal one interesting finding. The performance gain in RMSE is mostly from those users
              who indeed have concept drift in their user latent vectors at the time of prediction. In particular, for the synthetic dataset and the Ciao dataset, there are quite a few users with that property and the performance gains for these two datasets are more significant than those for the other datasets.

\end{itemize}

The rest of paper is organized as follows. In Section 2, we provide a review of related work. We define the rating prediction problem in Section 3 and propose the method of incorporating temporal dynamics including capturing and predicting the user preferences in Section 4. In Section 5, we conduct experiments on both synthetic and several real datasets  to validate our proposed temporal method. 
%In Section 6, we discuss the issues based on the observations in experiments. 
Finally, we conclude our work and point out the future research directions in Section 6.

\section{Related Work}
\label{c:review}

In this section, we first briefly review the MF approach for recommender systems and several recent approaches that intend to incorporate temporal dynamics with MF, including time-dependent collaborative filtering, tensor factorization, and collaborative Kalman filter.

\subsection{Matrix Factorization}

Matrix Factorization (MF)  performs well in the rating prediction task and has attracted considerable attention. The rationale behind the MF approach is to characterize each user and item by a series of latent factors that can be used for representing or approximating the interactions between users and items from the historical rating logs. Specifically, given an $M \times N$ rating matrix $R=(R_{i,j})$ with $M$ users and $N$ items, the MF approach considers the following optimization problem:
\begin{equation}\label{eq:2.1}
\min_{P,Q} \frac{1}{2} \sum_{i=1}^{M} \sum_{j=1}^{N} I_{ij} \left(R_{ij}-Q_{j}^{T} \cdot P_{i} \right)^{2} + \frac{\lambda}{2} \left(\left\| P \right\|^{2} + \left\| Q \right\|^{2} \right),
\end{equation}
where $P$ and $Q$ are the latent matrices which record the latent factors of users and items respectively.
Also, $P_i$ is the $i^{th}$ column of $P$, $Q_j$ is the $j^{th}$ column of $Q$, and $I_{ij}$  is an indicator function that is equal to $1$ if user $i$ rated item $j$ and equal to $0$ otherwise. The vector $P_i$, called the user latent vector of user $i$, is commonly used for representing (latent) user preferences, and the vector $Q_j$, called the user latent vector of item $j$, is commonly used for representing (latent) item characteristics.
The regularization terms are added in the optimization problem to prevent overfitting.

We can also view MF  from a probabilistic perspective. Probabilistic Matrix Factorization (PMF) \cite{mnih2007probabilistic, lawrence2009non} defines the following conditional distribution over the observed ratings based on the linear model with Gaussian observation noise:
\begin{equation}\label{eq:2.2}
p\left ( R|P,Q,\sigma^{2}\right) = \prod_{i=1}^{M}\prod_{j=1}^{N} \left[ N\left( R_{ij}|Q_{j}^{T} \cdot P_{i},\sigma^{2}\right) \right]^{I_{ij}},
\end{equation}
where $N\left( x|\mu,\sigma^{2}\right)$ denotes the probability density function of the Gaussian distribution with mean $\mu$ and variance $\sigma^{2}$. With placing zero-mean spherical Gaussian priors on the latent factors,  the problem of  maximizing the log-posterior with the fixed variance is equivalent to the optimization problem in \eqref{eq:2.1}.

Note that both $P_{i}$ and $Q_{j}$ are unknowns in  \eqref{eq:2.1} and the objective function is not {\em convex} \cite{koren2009matrix}. Simon Funk \cite{webb2006netflix} popularized a stochastic gradient descent (SGD) algorithm which loops through all ratings in the training set to find the  latent matrices $P$ and $Q$. For each training example $R_{ij}$, one first computes the associated prediction error
\begin{equation}\label{eq:2.3}
e_{ij} = R_{ij} - Q_{j}^{T} \cdot P_{i}.
\end{equation}
One then updates $P_{i}$ and $Q_{j}$ in the opposite direction of the gradient as follows:
\begin{align}
P_{i} &\leftarrow P_{i} + \alpha \left( e_{ij}Q_{j} -\lambda P_{i} \right), \label{eq:2.3a}\\
Q_{j} &\leftarrow Q_{j} + \alpha \left( e_{ij}P_{i} -\lambda Q_{j} \right), \label{eq:2.3b}
\end{align}
where $\alpha$ is the learning rate and $\lambda$ is the regulator parameter.
This incremental and iterative approach provides a practical way to scale the MF method to large datasets.

\subsection{Time-dependent Collaborative Filtering}

In order to provide recommendations that fit users' present preferences, time-dependent collaborative filtering (CF) \cite{shi2014collaborative} employs the availability of temporal information (time \break stamps) associated with user-item rating logs to put more emphasis on the recent ratings.
Such an approach is based on
the plausible assumption that recent logs have bigger influence on future events than old and obsolete logs.
There are many prior works on time-dependent collaborative filtering, including
neighborhood-based CF \cite{ding2005time,lathia2009temporal}, social influence analysis \cite{yang2014survey,goyal2010learning,palovics2014exploiting}, temporal bipartite projection \cite{wu2014temporal} and timeSVD++ \cite{koren2010collaborative}.
Among all these prior works, timeSVD++ \cite{koren2010collaborative} is perhaps the most related work to MF. In
  \cite{koren2010collaborative}, Koren  proposed adding a time-varying rating bias for each user and each item to the estimate from the original MF. As such, the temporal dynamics of user latent vectors are only modelled by a simple sum of three factors,
the stationary portion, a possible gradual change with linear equation of a deviation function, and a day-specific parameter for sudden drift.
Even so, it was reported in \cite{koren2010collaborative} that timeSVD++ significantly outperforms SVD and SVD++ \cite{koren2008factorization} (that considered implicit feedback).

Although these methods improve the accuracy of the prediction compared to the baseline MF estimator, there are some difficulties in the time-dependent CF approach. The system model in timeSVD++  for the user latent vectors is too simple to have any structural characterizations or constraints on their parameters. As such, these parameters (in various aspects and time steps) have to be learned individually and need lots of efforts on fine tuning.
  Thus, timeSVD++ maybe too data-specific to be used as a general model. Also, the assumption that claims recent ratings are always more important than  old data may be oversimplified.
%   and requires further parameter fitting to decide the size of time window and the time-decaying function.

\subsection{Tensor Factorization}

Tensor factorization (TF) extends MF into a three-dimen-sional tensor by incorporating the temporal features into the prediction model.  The underlying physical meaning of TF is that the given ratings not only depend on the user preferences and the item characteristics but also the current trend. There are two kinds of popular tensor factorization models in CF \cite{kolda2009tensor}: the CANDECOMP/PARAFAC (CP) model that decomposes the tensor into same rank of latent factors, and the Tucker model that considers the problem as the higher-order PCA.

There are some works that adopt the TF model for exploiting temporal information associated with user-item interactions. The Bayesian Probabilistic Tensor Factorization (BPTF) \cite{xiong2010temporal} extended PMF to CANDECOMP/PARAFAC tensor factorization that models each rating as the inner product of the latent factors of user, item, and time slice as well. It also imposes constraints that the adjacent time slices should share similar latent factors. The advantage of BPTF is its almost parameter-free probabilistic tensor factorization algorithm with a fully Bayesian treatment derivation while the drawback is it is not sensitive enough to capture the local changes of preferences compared with timeSVD++. Recently, Rafailidis and Nanopoulos \cite{rafailidis2014modeling} modeled continuous user-item interactions over time and defined a new measure of user preference dynamics to capture the shifting rate for each user. In a broader sense, recommendation can be regarded as a bipartite link prediction problem that aims to infer new interactions between users and items which are likely to occur in the near future. Based on this idea, Dunlavy et al. \cite{dunlavy2011temporal} considered bipartite graphs that evolve over time and demonstrated that tensor-based methods are effective for temporal data with varying periodic patterns. Apart from incorporating the temporal information, tensor factorization is a popular approach to integrate further information such as context of implicit feedback in content-based recommender systems. For instance, Moghaddam et al. \cite{moghaddam2012etf} added {\em review} as the third dimension based on the Tucker tensor model to address the problem of personalized review quality prediction and Shi et al. \cite{shi2012tfmap} directly trained the tensor model for creating an optimally ranked list of items for individual users in the context-aware recommender systems.

Tensor factorization provides a principled and well-struc-tured approach to incorporate the temporal dynamics in recommender systems; however, the structure also limits the flexibility of the model so that it is hard to process and solve the decomposition especially for a large-scale and sparse tensor. Given the same amount of rating data, the higher order the tensor model is, the more severe the sparsity problem is. The sparsity problem leads to time-consuming computing, high space complexity and the convergence issues in the decomposition procedure.
%In our temporal dynamic model, we will try to overcome this challenge by introducing some overlaps between the time %steps.

\subsection{Collaborative Kalman Filter}

Inspired by the success of PMF that places Gaussian priors on the latent factors and formulates the matrix factorization problem as an optimization problem for obtaining the Maximum-a-Posteriori (MAP) estimate, there are some recent works that compute the MAP optimally by using the Kalman filter
\cite{kalman1960new}. Considering the observed measurements over time with noise and uncertainties, the Kalman filter is the optimal linear estimator of unknown variables. Its recursive structure also allows new measurements to be processed as they arrive. The Kalman filter can be conceptualized in two phases: the {\em predict} phase is called a priori estimate which produces an estimation without the observation at the current time step, and the {\em update} phase is known as a posteriori estimate which refines the estimation with the current observation. The refinements of the state and covariance estimates are based on the optimal Kalman gain computed at every time step.

   The  paper \cite{lu2009spatio} by
Lu et al. might be the first paper to use the Kalman filter in recommender systems. In that paper, they exploited the Kalman filter to model the change of user preferences in its temporal component. Though they provided a new perspective to the recommender systems, their approach is still not
general enough  as
the transition matrix used in the Kalman filter was only modeled by an {\em identity} matrix. As such, one can only capture the drifts of user preferences.
In another recent paper \cite{gultekin2014collaborative}, Gultekin and Paisley proposed the collaborative Kalman filter
(CKF) approach that used a geometric Brownian motion to model the dynamically evolving drift of each latent factor.
 The dynamic state space model proposed by \cite{sun2012dynamic, sun2014collaborative} is most related to our work. To solve the system identification problem for the linear system in the Kalman filter, they develop an EM algorithm that performs the Kalman filter and the RTS smoother. The EM algorithm is an iterative two-pass algorithm that yields  estimates for the model parameters by using all observations in the expectation step, and then refines the estimates of the model parameters in the maximization step.
Although the model is comprehensive and provides better results compared to the SVD and timeSVD++ approaches,
  there are some limitations in practice: (i) it makes a very strong assumption that assumes the transition matrix is homogeneous for all users. Such a homogeneous assumption is needed to simplify the model (as otherwise it is very difficult, if not impossible, to determine all their parameters form the EM algorithm \cite{sun2014collaborative}), and (ii) it is not suitable for large datasets due to the tractability and runtime performance.

In this paper, we will remove the assumption that the transition matrix is homogeneous for all users. By doing so, we allow our system to track
concept drift in each individual user latent vector. Our experiments further verify that users do have different transition matrices. Some of them are simply governed by the identity matrix and have no concept drift in their latent vectors. On the other hand, some of them have significant changes of their latent vectors and the improvement of the rating for those users is the key factor for lowering RMSE in our temporal approach.

%To sum up, the main reason to these obstacles is without the understanding of the whole system when applying Kalman filter to the recommendation %problem. We define our temporal approach with the help of the Kalman filter model while propose different principled methods to capture the temporal %dynamics and learn the transition matrix at the individual level to better understand the evolution of each user for further improving the performance %of recommendation.

\section{Problem Definition}
\label{c:definition}

In this paper, we study the rating prediction problem with time stamped logs. Specifically, there are $M$ users, indexed from $i=1,2, \ldots$, $M$, and $N$ items, indexed from $j=1,2, \ldots, N$. For these users and items, we are given a set of time stamped
logs, where each log is represented by the four tuple:
$$\mbox{(user, item, rating, time)}.$$
We assume that every rating is a real-valued number and each item can be rated by a user at most once.
If we neglect the time stamps of these logs, then the ratings of these logs can be represented by an $M \times N$ matrix $R=(R_{ij})$, where $R_{ij}$ is the rating of user $i$ on item $j$ if item $j$ has been rated by user $i$. On the other hand, if item $j$ has not been rated by user $i$, then $R_{ij}$ is said to be {\em missing}.
In practice, the matrix $R$ is a {\em sparse} matrix and there are many missing values.
The rating prediction problem is then to predict the missing values in the matrix $R$.

To evaluate the performance of a rating prediction algorithm, the rating logs are partitioned into two sets: the training set and the testing set. The training set is given to a rating prediction algorithm to ``learn'' the needed parameters for rating prediction. On the other hand, the testing set is not revealed to a rating prediction algorithm and is only available for testing the accuracy of a rating prediction algorithm. Though there are many  metrics  for evaluating
the performance of rating prediction algorithms, in this paper we adopt the root mean square error (RMSE) that can be computed as follows:
\begin{equation}
\label{eq:RMSE}
RMSE = \sqrt{\frac{\sum_{\left ( i,j \right )\in Testing\: Set}\left ( R_{ij}- \hat{R_{ij}} \right )^{2}}{\left | Testing\: Set \right |}},
\end{equation}
where $\hat R_{ij}$ is the prediction for $R_{ij}$ via a rating prediction algorithm.
The RMSE has been widely used in the literature, including the competition for the Netflix Prize.
Although the range of RMSE might be quite small, there is evidence (see e.g., \cite{koren2008factorization}) that small improvement in RMSE can have a significant impact on the quality of the top few recommendations from a rating prediction algorithm.

The original MF does not use the information of time stamps and simply decomposes the matrix $R$ approximately into a product of two matrices: the user latent matrix and the item latent matrix. Though the item latent matrix could be quite stationary with respect to time, it is a general believe
(see e.g., \cite{koren2010collaborative,lu2009spatio,sun2012dynamic,sun2014collaborative}) there might be concept drift in the user latent matrix as users tend to change their mind over time. In view of this, our aim is to develop a temporal dynamic model for tracking concept drift in each individual user latent vector by using the time stamps of rating logs. By doing so, we can  effectively and efficiently achieves a lower RMSE than that of MF.

\section{Temporal Matrix Factorization}
\label{c:method}

For the rating prediction problem described in the previous section, we propose a Temporal Matrix Factorization (TMF) approach that is capable of tracking
concept drift in each individual user latent vector. Our approach is based on the following assumptions that were previously used in
the literature (see e.g., \cite{koren2010collaborative,lu2009spatio,sun2012dynamic,sun2014collaborative}):
\begin{description}
\item[(i)]
Like the original MF, there are a  user latent matrix $P  \in {\cal R}^{D\times M}$ and an  item latent matrix $Q \in {\cal R}^{D\times N}$
that can be used for approximating the rating matrix $R$. The $i^{th}$ column of the user latent matrix $P$, denoted by $P_i$, is called the user latent vector of user $i$, $i=1,2, \ldots, M$, that can be viewed as the preferences of user $i$ for the $D$ latent factors. Similarly, the $j^{th}$ column vector of the item latent matrix $Q$, denoted by $Q_j$, is called the item latent vector
of item $j$, $j=1,2, \ldots, N$. The rating for user $i$ on item $j$ is then predicted by the inner product of $P_i$ and $Q_j$.
\item[(ii)] There is concept drift in each individual user latent vector as people might change their preferences over time. For this, we denote by $P(t)$ the user latent matrix at time $t$, and $P_i(t)$ the user latent vector of user $i$ at time $t$, $i=1,2, \ldots, M$.
\item[(iii)] As the characteristics of items are stationary,  we assume that the item latent matrix $Q$ is invariant with respect to time.
\end{description}

In view of  these assumptions, the key ingredient of  our TMF approach is to use the training data set to capture the dynamics of the concept drift in each individual user latent vector. For this, our approach consists of the following steps:
\begin{description}
\item[(i)] Use the rating logs in the training data set to construct a time series of $M \times N$ rating matrices, $\{R(t), t =1,2, \ldots, T-1\}$.
\item[(ii)] Use the time series of rating matrices $\{R(t), t =1,2, \ldots$, $T-1\}$ to learn a time series of $D \times 1$  user latent vectors, $\{P_i(t), t =1,2, \ldots, T-1\}$, $i=1,2, \ldots, M$.
\item[(iii)] For each user $i$, use the  time series of  user latent vectors, $\{P_i(t), t =1,2, \ldots, T-1\}$ to learn the  dynamics of the concept drift in the user latent vector.
\item[(iv)] Use the dynamics of the concept drift in each individual user latent vector to predict the user latent vector at time $T$, i.e., $P(T)$. Then use the  product of $P(T)$ and the item latent matrix $Q$ to predict the missing values in the testing data set.
\end{description}

\subsection{Construction of a time series of rating matrices}
\label{sec:rating matrices}

The simplest way to construct a time series of  rating matrices $\{R(t), t =1,2, \ldots, T-1\}$ is to partition rating logs into equally spaced time slices according to their time stamps. But, as the original rating matrix in a real data set might have already been very sparse, further partitioning of the rating logs might yield a time series of extremely sparse rating matrices that might not have any statistical significance at all. In view of the sparsity problem, the number of time slices $T$ cannot be too large.
To further mitigate the sparsity problem, one can consider a sliding window approach that merges the rating logs in several consecutive time slices into a single step. By doing so, there are {\em overlapping} rating logs in such a time series of  rating matrices. Such an approach can not only mitigate the sparsity problem but also ensure smooth change of  rating matrices so that prediction could be possible.

\subsection{Learning a time series of user latent vectors}
\label{sec:latent matrices}

To learn a time series of user latent vectors for each user, we first perform MF for the rating matrix $R$ to obtain the
user latent matrix $P$ and the item latent matrix $Q$.
As we assume that the item latent matrix $Q$ is invariant with respect to time, one might expect that
\begin{equation}\label{eq:4.1}
R_{i}(t) = Q^T \cdot P_{i}(t),
\end{equation}
where $R_i(t)$ is the rating vector for user $i$ on the $N$ items (that can be extracted from the rating matrix $R(t)$).
In view of this,  a n\"aive way to
learn a time series of $D \times 1$  user latent vectors, $\{P_i(t), t =1,2, \ldots, T-1\}$, is to simply compute
the  Moore-Penrose pseudoinverse of $Q^T$ from (\ref{eq:4.1}).
It is well-known that the Moore-Penrose pseudo inverse computes a ``best fit,'' i.e., the least squared solution to a system of linear equations and its uniqueness follows from the SVD theorem in matrix algebra. Such an approach works fine if all the entries in the vector $R_i(t)$ are known. In reality, there are many missing values in the vector $R_i(t)$ and thus make a direct computation of the Moore-Penrose pseudo inverse infeasible.
One way to remedy this is to pad the missing values in $R_i(t)$ with the predicted values of user $i$ by MF, i.e., $Q^T \cdot P_i$. In particular, one can generate another vector $\tilde R_i(t)$ as  a linear combination of these two vectors, i.e.,
\begin{equation}\label{eq:4.2c}
\tilde R_i(t)=\beta R_i(t)+(1-\beta)Q^T \cdot  P_i.
\end{equation}
If $\beta$ is small,  the padded values in $\tilde R_i(t)$ are all from the vector $Q^T \cdot P_i$. As such, the vectors $\tilde R_i(t), t =1,2, \ldots, T-1$,
are all very similar and the corresponding Moore-Penrose pseudo inverse vectors also very similar. As a result, there is basically no change of the user latent vectors and that defeats the purpose of tracking the dynamics of user latent vectors.
On the other hand, if $\beta$ is large, then we basically ignore all the missing values in $R_i(t)$ and
that causes great fluctuation of the user latent vectors which makes it extremely difficult to track the dynamics of user latent vectors.
Also, as there are many missing values in $R_i(t)$, it is not clear whether the user latent vectors obtained this way possess any statistical significance.

The key insight to tackle this problem is that the user preferences at a specific time step are not only related to  the ratings during  that specific time step but also related to his/her overall behavior. In view of this, we first set $P_{i}(t)$  as the original user latent vector $P_{i}$. Then we use the observed ratings during that time step to ``learn'' $P_i(t)$.
%Specifically, we consider the following MF type of objective function:
%\begin{equation}\label{eq:4.3}
%\min_{P_{i}(t)} \sum_{j=1}^{N} I_{ij}(t) \left(R_{ij}(t)-Q_{j}^{T} \cdot P_{i}(t) \right)^{2} + \frac{\lambda}{2} %\cdot \left\| P_{i}(t) \right\|^{2}.
%\end{equation}
Specifically, we propose the following modified stochastic gradient descent method:
\begin{align}
e_{ij}(t) &= R_{ij}(t) - Q_{j}^{T} \cdot P_{i}(t), \label{eq:4.4}\\
P_{i}(t) &\leftarrow P_{i}(t) + \alpha \left[ e_{ij}(t)Q_{j} -\lambda P_{i}(t) \right].\label{eq:4.5}
\end{align}
Unlike the standard stochastic gradient descent method for MF in (\ref{eq:2.3})--(\ref{eq:2.3b}),
 here we only update the latent vector $P_{i}(t)$ for every rating provided by user $i$ at time $t$ (as the  item matrix $Q$ is stationary).
By doing so, the user latent vector $P_i(t)$ only updates his/her preferences for those items rated during time $t$ and thus retains his/her overall behavior for those items not rated during time $t$. Such an approach not only overcomes the obstacle of data sparsity but also possesses meaningful user preferences in the temporal setting.

\subsection{Learning the  dynamics of the concept drift in the user latent vector}
\label{sec:dynamic}

To track concept drift in the user latent vector, we need to identify a system model for the dynamics of the time series of the user latent vectors. Such a problem is known as the {\em system identification} problem in the literature \cite{johansson1993system}. One of the most commonly used models for system identification problems is the linear system model. As such,
 we consider the linear system model  for the latent vector  of each user.
 Specifically, we consider $P_i(t)$ as the state vector at time $t$ and model the evolution of the state vector by
\begin{equation}\label{eq:4.2}
P_{i}(t) = A_{i} \cdot P_{i}(t-1) + b_{i},
\end{equation}
where $A_i$ is a $D \times D$ matrix and $b_i$ is a $D \times 1$ vector.
The matrix $A_i$  is called the {\em transition} matrix for user $i$ and $b_i$ is called the  {\em bias} vector of user $i$.

It seems plausible to assume that
 the user latent vectors do not vary a lot in each time step. As such, we replace the transition matrix $A_{i}$ by $\left( I+\tilde{A_{i}} \right)$ in (\ref{eq:4.2}). This then leads to
\begin{equation}
Z_{i}(t) =\tilde  A_{i} \cdot P_{i}(t-1) + b_{i}, \label{eq:4.10}
\end{equation}
where
\begin{equation}\label{eq:4.9}
Z_{i}(t) = P_{i}(t) - P_{i}(t-1).
\end{equation}
By doing so, we expect that the
matrix $\tilde{A_{i}}$ is sparse and it only contains  a small number of nonzero entries.
It is known \cite{tibshirani1996regression} that the Lasso regression provides parameter shrinkage and variable selection that limit the number of nonzero elements in the parameters. As such, we apply the Lasso regression to estimate $\tilde A_i$ and $b_i$ in (\ref{eq:4.10}) from the $T-2$ ``observations'' of the output $\{Z_i(t), t=2, \ldots, T-1\}$ with the input $\{P_i(t), t=2, \ldots, T-1\}$. Specifically, for each factor $k$, $k=1,2, \ldots, D$, we let
$A_i^{(k)}$ be the $k^{th}$ row of $A_i$, $b_i^{(k)}$ be the $k^{th}$ element in $b_i$, $Z_i^{(k)}(t)$ be the
$k^{th}$ element in $Z_i(t)$ and consider
the following optimization problem:
\begin{eqnarray}\label{eq:4.12}
&&\min_{\tilde A_{i}^{(k)}, b_{i}^{(k)}} {1 \over {2(T-2)}}\sum_{t=2}^{T-1}\left (Z_{i}^{(k)}(t)-\tilde A_{i}^{(k)} \cdot P_i(t) -b_{i}^{(k)} \right )^{2} \nonumber\\
&&\quad\quad\quad\quad +\lambda \|\tilde{A_{i}}^{(k)} \|_1,
\end{eqnarray}
where $\|\tilde{A_{i}}^{(k)} \|_1$ is the $L_1$-norm of the vector $\tilde{A_{i}}^{(k)} $ and
$\lambda$  is a nonnegative regulator parameter for the Lasso regression. As $\lambda$ increases, the number of nonzero elements in the vector $\tilde{A_{i}}^{(k)}$ decreases.
In our experiments, we will use the Matlab tool \cite{matlabLasso} to solve the above Lasso regression.

\subsection{Rating prediction}
\label{sec:prediction}

Once we obtain the transition matrix $A_{i}$ and the bias vector $b_i$, we can use the system dynamic in (\ref{eq:4.2}) to predict the latent vector of user $i$ at time $T$ by the following equation:
\begin{equation}
P_{i}(T) = A_{i}\cdot  P_{i}(T-1) + b_{i}. \label{eq:4.13}
\end{equation}
As in the original MF, the missing values  in the testing data set are then predicted by using the  product of the user latent vector and the item latent matrix, i.e.,
\begin{equation}
R_{i}(T) = Q^T  \cdot P_{i}(T). \label{eq:4.14}
\end{equation}

\section{Experiment Results}
\label{c:experiments}
%\begin{spacing}{2}

In this section, we perform various experiments to evaluate the performance and efficiency of our temporal method via a synthetic dataset and four real datasets. All our experiments are implemented in MATLAB and executed on a server equipped with an Intel Core i7 (4.2GHz) processor and 64G memory on the Linux system.

\subsection{Experiments on the Synthetic Dataset}

We first conduct our experiments on a synthetic data. The main reason of doing this is to test our method in a {\em controllable} environment so that we can gain insights of the effects of various parameters and thus better understand when our method could be effective.

 To generate the synthetic data, we set $M=10,000$ and $N=10,000$, i.e., there 10,000 users and 10,000 items. The density of the rating matrix $R$ is set to 1\%, and that
  gives 1,000,000 ratings.  We generate all the entries in both the initial user latent matrix $P(1)$ and the item latent matrix $Q$ by uniformly distributed random variables over $(0,1)$. To model the evolution of the user latent vector for each user $i$, the transition matrix $A_{i}$ is generated by the sum of the identity matrix and a  random matrix $R^\prime$ with all its entries generated from a uniform distribution. The entries in the bias vector $b_{i}$ are  also generated from a uniform distribution. Various ranges of the entries in $R^\prime$ and $b_i$ are specified in our experiments (see Table \ref{table:table1}). The number of steps $T$ is set to 10 and the  rating logs are then generated according to equations \eqref{eq:4.1} and \eqref{eq:4.2}.

In Table \ref{table:table1}, we report the RMSE for both the original MF (implemented by the LIBMF library \cite{chin2015fast, chin2015learning}) and our method  for various parameter settings. The parameters that we choose for MF are the learning rate $\alpha$= 0.01, the regulator parameter  $\lambda= 0.02$, and the number of factors $D = 30$. We run 50 iterations of SGD to obtain the  latent factors in the original MF. The learning rate $\alpha$ and the regulator parameter $\lambda$ in equation \eqref{eq:4.5} for computing the user latent vector at every time step in our method are set to be the same as those for MF. As can be seen from this table,  our method consistently and significantly outperforms MF in all the parameter settings. The improvement depends on the transition matrix and the bias vector that are selected to control the concept drift in the user latent vector. A more careful examination reveals that the improvement for our method is relatively small if the  range of the entries in the random matrix $R'$ is small, e.g., $(-0.01, 0.01)$. This is because the transition matrix in such a scenario is very close to the identity matrix and there is almost no change of the user latent vectors. As such, MF performs well and yields a low RMSE.  On the other hand,  if such a range is large,
the prediction accuracy of MF is low as MF relies on the assumption of stationary user preferences. As our method is capable of tracking the concept drift in the user latent vector, our method achieves roughly 17-26\% improvement in terms of RMSE.

Next, we study the effect of the range of $b_{i}$. In Table \ref{table:table1}, we consider two different ranges of $b_i$. The experimental results show that the values of RMSE are larger when the given range is larger (under the condition of using the same transition matrix). Moreover, the performance gain from our  method is also larger. This shows the importance of adding the bias vectors  in our linear system model.

%Overall, the results in this synthetic data set demonstrate that our method is able to deal with the time-changing rating data and further make more %accurate predictions. It breaks the limitation of the original MF approach that only measure the static properties.

\begin{table}
{\tiny
\begin{center}
\caption{The RMSE results on the synthetic dataset for various parameter settings.}
\begin{tabular}{|c|c|c|c|c|c|}
\hline
Range of $R'$ & Range of $b_{i}$ & MF & Our method & Improvement \\ \hline\hline
(-0.01, 0.01) & \multirow{5}{*}{(-0.01, 0.01)} & 0.4566 & 0.4246 & 7.01\% \\ \hhline{-~---}
(-0.05, 0.05) & & 0.8468 & 0.6758 & 20.19\% \\ \hhline{-~---}
(-0.1, 0.1) & & 1.2667 & 0.9790 & 22.71\% \\ \hhline{-~---}
(-0.3, 0.3) & & 1.7007 & 1.3780 & 18.97\% \\ \hhline{-~---}
(-0.5, 0.5) & & 1.8046 & 1.4967 & 17.06\% \\ \hline\hline
(-0.01, 0.01) & \multirow{5}{*}{(-0.1, 0.1)} & 0.5763 & 0.5124 & 11.09\% \\ \hhline{-~---}
(-0.05, 0.05) & & 0.9645 & 0.7660 & 21.20\% \\ \hhline{-~---}
(-0.1, 0.1) & & 1.2985 & 0.9539 & 26.54\% \\ \hhline{-~---}
(-0.3, 0.3) & & 1.7092 & 1.3673 & 19.96\% \\ \hhline{-~---}
(-0.5, 0.5) & & 1.8091 & 1.4787 & 18.26\% \\ \hline
\end{tabular}
\label{table:table1}
\end{center}
}
\end{table}

\begin{figure}
\begin{center}
\includegraphics[width=0.7\columnwidth]{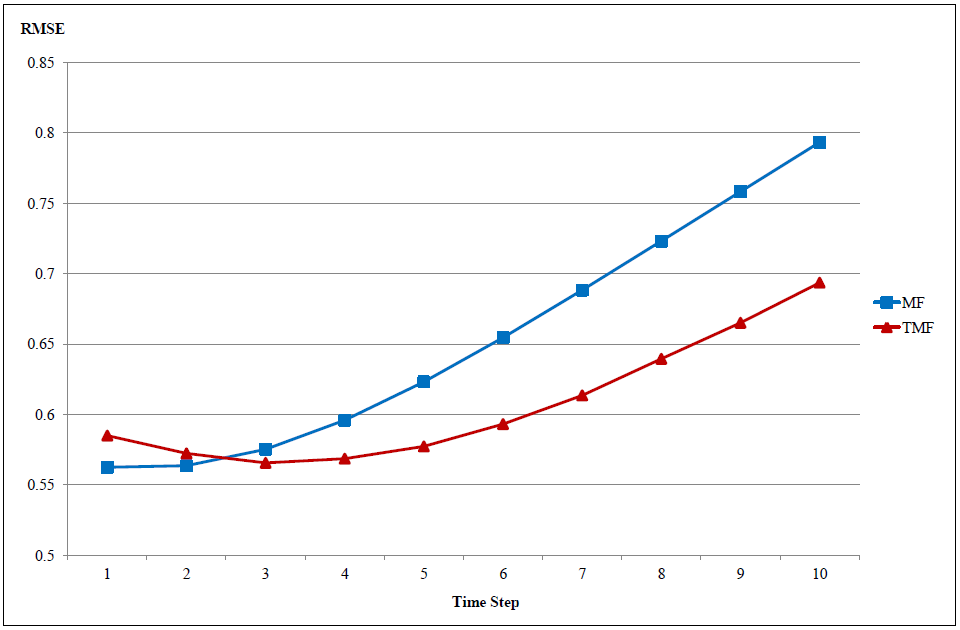}
\caption{The average dissimilarities among all users at each time step.}
\label{fig:1}
\end{center}
\end{figure}

In addition to the prediction results presented above, we demonstrate the state tracking ability of our temporal \break method. Denote by $\hat{P}_{i}(t)$ the user latent vector of user $i$ at time $t$  (computed by our temporal method) and $P_{i}(t)$ the given ground truth in the synthetic dataset. We compute the dissimilarity measure of these two latent vectors by using the RMSE metric as follows:
\begin{equation}\label{eq:5.1}
s\left(P_{i}(t), \hat{P}_{i}(t) \right) = \sqrt{\frac{\sum_{d=1}^{D}\left ( P_{id}(t)- \hat{P_{id}}(t) \right )^{2}}{D}},
\end{equation}
where $D$ is the dimensions of these two latent vectors.
To compare the tractability of the MF approach and our temporal method, we measure the average of the dissimilarities among all the users at a specific time step $t$ and plot the results in Figure \ref{fig:1}.  As can be seen from Figure \ref{fig:1}, the gain of using our temporal method to track the user latent vectors increases over time and at the time for prediction, i.e., the $10^{th}$ time step, the gain is near 13\%.

To dig further, we examine the rating prediction result for each user.
 We observe that our temporal method consistently obtains better prediction results if the MF approach always overestimates (or underestimates)  all the ratings provided by a user in the testing dataset. For instance, we show in Table \ref{table:table2} a user (user ID 26) who gives extreme high ratings and another user (user ID 28) who  gives extremely low ratings. This is because  the  MF approach cannot track the  concept drift in the user latent vector even when
 there is a distinct downward (or upward) trend in the evolution of user preferences. On the other hand, if
 the ratings provided by a user are distributed over the entire range as shown in Table \ref{table:table3}, then
 the ratings predicted by our temporal method are not always better than those from those predicted by the MF.
 But, the overall accuracy of our temporal method is still better.
  In summary, our temporal method is good at capturing the general trend and thus yields improvement on the overall performance.

\begin{table}
{\tiny
\begin{center}
\caption{The rating prediction results with extreme actual ratings.}
\begin{tabular}{|c|c|c|c|c|}
\hline
user ID & item ID & actual rating & MF  & Our method \\ \hline\hline
\multirow{10}{*}{26} & 1139 & 5 & 4.51  & 4.95 \\ \hhline{~----}
& 1303 & 5 & 4.62  & 5.07 \\ \hhline{~----}
& 2092 & 5 & 3.87  & 4.26 \\ \hhline{~----}
& 2200 & 5 & 4.22  & 4.63 \\ \hhline{~----}
& 2625 & 5 & 3.83  & 4.21 \\ \hhline{~----}
& 2867 & 5 & 4.43  & 4.86 \\ \hhline{~----}
& 3515 & 5 & 4.69  & 5.14 \\ \hhline{~----}
& 6495 & 5 & 4.67  & 5.13 \\ \hhline{~----}
& 7864 & 5 & 3.71  & 4.07 \\ \hhline{~----}
& 8693 & 5 & 4.21  & 4.62 \\ \hline\hline
\multirow{10}{*}{28} & 92   & 1 & 2.24  & 0.88 \\ \hhline{~----}
& 1440 & 1 & 2.44  & 0.96 \\ \hhline{~----}
& 1626 & 1 & 3.08  & 1.18 \\ \hhline{~----}
& 1917 & 1 & 3.15  & 1.27 \\ \hhline{~----}
& 3234 & 1 & 2.95  & 1.15 \\ \hhline{~----}
& 3556 & 1 & 2.62  & 0.98 \\ \hhline{~----}
& 4425 & 1 & 2.91  & 1.10 \\ \hhline{~----}
& 8990 & 1 & 2.23  & 0.85 \\ \hhline{~----}
& 9262 & 1 & 2.37  & 0.94 \\ \hhline{~----}
& 9978 & 1 & 2.79  & 1.09 \\ \hline
\end{tabular}
\label{table:table2}
\end{center}
}
\end{table}

\begin{table}
{\tiny
\begin{center}
\caption{The rating prediction results when actual ratings are distributed over the entire range.}
\begin{tabular}{|c|c|c|c|c|}
\hline
user ID & item ID & actual rating & MF  & Our method \\ \hline\hline
\multirow{10}{*}{32} & 976  & 3 & 2.74  & 3.09 \\ \hhline{~----}
& 1224 & 4 & 2.63  & 2.98 \\ \hhline{~----}
& 1379 & 5 & 3.17  & 3.64 \\ \hhline{~----}
& 1691 & 2 & 2.72  & 3.06 \\ \hhline{~----}
& 3989 & 4 & 2.86  & 3.25 \\ \hhline{~----}
& 5079 & 1 & 2.62  & 2.95 \\ \hhline{~----}
& 6541 & 2 & 2.60  & 2.95 \\ \hhline{~----}
& 7455 & 4 & 2.58  & 2.93 \\ \hhline{~----}
& 9691 & 3 & 2.46  & 2.79 \\ \hhline{~----}
& 9944 & 3 & 2.47  & 2.81 \\ \hline
\end{tabular}
\label{table:table3}
\end{center}
}
\end{table}

\subsection{Experiments on the Real Datasets}

To validate our temporal method in practical environments, in this section we conduct our experiments on   real datasets. Motivated by the increasing importance of recommender systems on electronic commerce and online video streaming services, we consider  Ciao, Epinions, Flixster and MovieLens. With Web 2.0 technique to gather users' feedbacks such as explicit ratings and implicit reviews, Ciao and Epinions are two of the most popular online-shopping websites. Ciao is a European-based online-shopping portal with websites and claims and it reaches an audience of 28.4 million monthly unique visitors in Europe \cite{CiaoWiki}. Epinions is established in 1999 and now is the largest consumer review site with thousands of product reports in the world. Another focus is the movie/video recommendation platform which becomes more popular in our daily life. For this, we consider Flixster and MovieLens in our work. Flixster is an American social movie site which also provides applications in Facebook and MySpace for users to share film reviews and ratings whereas MovieLens is a recommender system for research of collaborative filtering run by GroupLens Research. These datasets with the information of ratings and the associated time stamps are publicly available in \cite{CiaoEpinions, Flixster, MovieLens} and their statistics information is shown in Table \ref{table:table4}.

\begin{table}
{\tiny
\begin{center}
\caption{Statistics of the real datasets.}
\begin{tabular}{|r|c|c|c|c|}
\hline
& Ciao & Epinions & Flixster & MovieLens \\ \hline\hline
Users & 1,947 & 21,752 & 114,747 & 125,041\\ \hline
Items & 5,004 & 242,842 & 44,439 & 17,951\\ \hline
Training Ratings & 22,068 & 830,043 & 7,376,472 & 18,000,243\\ \hline
Testing Ratings & 826 & 23,621 & 416,293 & 353,575\\ \hline
 Density & 0.23\% & 0.02\% & 0.14\% & 0.80\%\\ \hline
Earliest Rating & Jun. 2000 & Jul. 1999 & Dec. 2005 & Jan. 1995\\ \hline
Latest Rating  & Apr. 2011 & May 2011 & Nov. 2009 & Mar. 2015\\ \hline
\end{tabular}
\label{table:table4}
\end{center}
}
\end{table}

\iffalse
\begin{table}
\begin{center}
\begin{tabular}{|r|c|c|c|c|}
\hline
& Ciao & Epinions & Flixster & MovieLens \\ \hline\hline
Users & 1,947 & 21,752 & 114,747 & 125,041\\ \hline
Items & 5,004 & 242,842 & 44,439 & 17,951\\ \hline
Training Ratings & 22,068 & 830,043 & 7,376,472 & 18,000,243\\ \hline
Testing Ratings & 826 & 23,621 & 416,293 & 353,575\\ \hline
Data Density & 0.23\% & 0.02\% & 0.14\% & 0.80\%\\ \hline
Earliest Rating & Jun. 2000 & Jul. 1999 & Dec. 2005 & Jan. 1995\\ \hline
Latest Rating  & Apr. 2011 & May 2011 & Nov. 2009 & Mar. 2015\\ \hline
\end{tabular}
\caption{Statistics of the real datasets.}
\label{table:table4}
\end{center}
\end{table}
\fi

All of these platforms provide services for users to rate items using a 5-point Likert scale while Flixster adopts 10 discrete numbers in the range [0.5,5] with step size 0.5. Each log in a dataset contains the information of user ID, item ID, rating and timestamp. First of all, we sort these logs  in the chronological order to form  a time series. Note that this setting is more practical than the traditional approach because we are only allowed to use the past data to predict future events. We partition the whole dataset into 10 time slices equally and leave the last slice as the testing set.
 In order to have an enough number of representative ratings and have smooth and trackable transitions, we apply the sliding window approach to combine the logs in every 5  consecutive time slices to form a time step. By doing so, 4 of the slices in each step overlap with those in the next time step and there are totally 6 time steps ($T$ = 6) in this setting. Next, we remove new users and new items, i.e., the users and items appear only once in the testing set, and focus on tracking the  evolution for the latent vectors of existing users. We adopt the same parameter settings as those in the synthetic dataset except that we choose the learning rate $\alpha$ = 0.005 (in \eqref{eq:4.5}) for the Flixster and MovieLens datasets. The experimental results for RMSE by the MF method and our method are shown in Table \ref{table:table5}.
\begin{table}
{\tiny
\begin{center}
\caption{The RMSE results for the four real datasets.}
\begin{tabular}{|c|c|c|c|c|}
\hline
& Ciao & Epinions & Flixster & MovieLens \\ \hline\hline
MF & 1.1099 & 1.1287 & 1.1189 & 0.8170\\ \hline
Our method & 1.0540 & 1.1189 & 1.1102 & 0.8150\\ \hline
Improvement & 5.04\%  & 0.87\% & 0.78\% & 0.24\%\\ \hline
\end{tabular}
\label{table:table5}
\end{center}
}
\end{table}

In comparison with the MF approach, we can see from Table \ref{table:table5} that our temporal method improves the performance in RMSE  in all four real datasets. However, the gain varies from one to another. In Ciao, we obtain 5\% improvement.  For the other three datasets, the improvements are not that significant. One possible explanation for this is that the temporal effect depends on the dataset due to its intrinsic properties. We also observe the improvements in real datasets are not as significant  as those in the synthetic dataset. The main reason is that we purposely construct the synthetic dataset so that it possesses the desired concept drift in the user latent vector.
It is not clear whether there are concept drifts in the user latent vectors in the Epinions, Flixster, and MovieLens datasets.

\begin{figure*}[!t]
\centering
%\subfloat[The number of channels is 13.]
{\includegraphics[width=0.6\columnwidth]{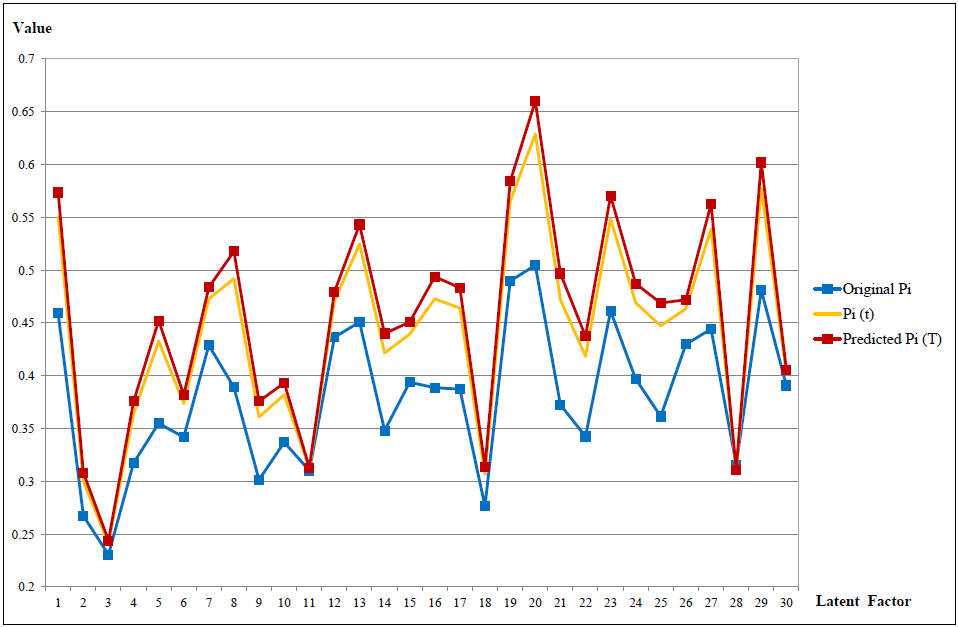}%
%\label{fig_first_case}
}
\hfil
%\subfloat[The number of channels is 23.]
{\includegraphics[width=0.6\columnwidth]{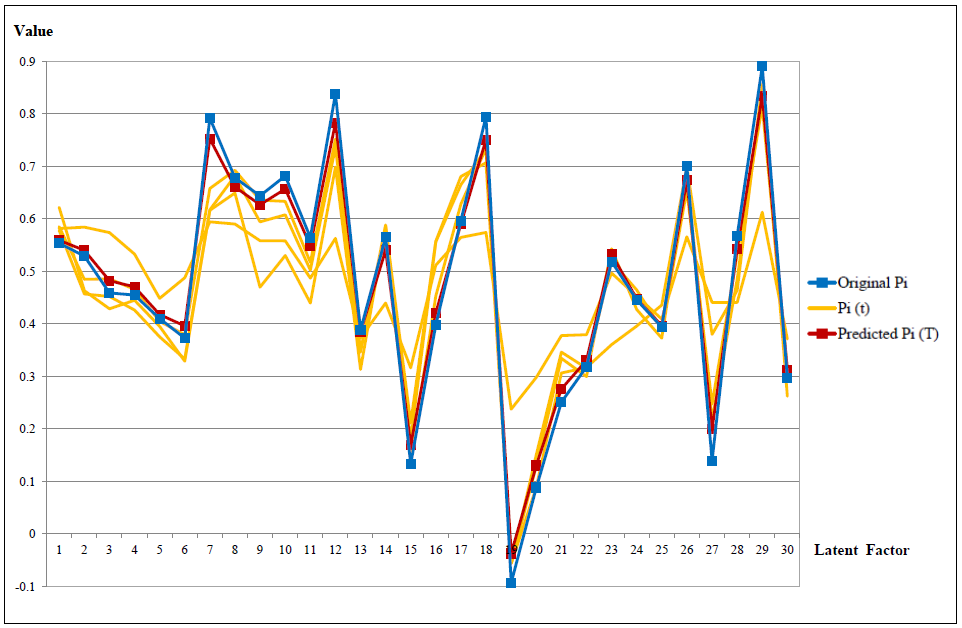}%
%\label{fig_second_case}
}
\hfil
%\subfloat[The number of channels is 23.]
{\includegraphics[width=0.6\columnwidth]{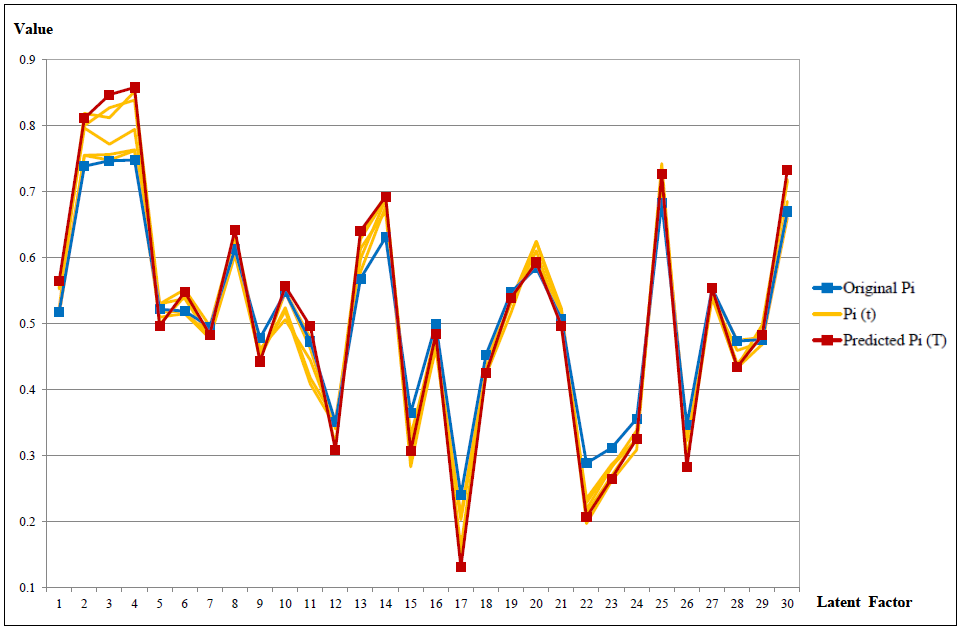}%
%\label{fig_third_case}
}
\caption{The evolution of various user latent vectors (from left to right):  user 49 in the Ciao dataset,  user 108 in the Ciao dataset, and user 339 in the Epinions dataset.}
\label{fig:2}
\end{figure*}

To see whether there is performance gain by tracking concept drift in the latent vector of a user,
we further examine the evolution of various user latent vectors and their corresponding rating prediction results (see Figure \ref{fig:2}).
% (see Figure \ref{fig:2}--Figure \ref{fig:4}).
An interesting finding is that users basically can be classified into two types: one is {\em beneficial} to track concept drift in his/her user latent vector and the predicted latent vector from our method is substantially different from that from MF,  and the other is {\em worthless} to
 track concept drift in his/her user latent vector as the predicted latent vector from our method is quite close to that from MF.
 In the Ciao dataset, we observe that the majority of improvement made by the users whose latent vectors evolve in a consistent direction. A plain example of this case is that the user latent vector changes in only one step from the original latent vector. We list the evolution of the first five latent factors of the latent vector (due to space limitation) and the corresponding ratings in Table \ref{table:table6}. To visualize this case more clearly, we also plot the factors of the latent vector for the original $P_{i}$ (computed by MF and marked in blue), the factors of the user latent vector at time step $T-1$ (marked in yellow), and the factors of the predicted user latent vector $P_{i}(T)$ with $T=6$ (marked in red) for user 49 in the Ciao dataset in the leftmost figure in Figure \ref{fig:2}. As can be seen from this figure,
  the predicted latent vector from our method is substantially different from that from MF. For this user, the corresponding ratings in Table \ref{table:table6} show that such a user is beneficial to track concept drift in his/her user latent vector.
On the other hand, we plot in the middle figure of Figure \ref{fig:2} the factors of the latent vector for the original $P_{i}$ (computed by MF and marked in blue), the factors of the user latent vector at time step $t$ (marked in yellow), and the factors of the predicted user latent vector $P_{i}(T)$ with $T=6$ (marked in red) for user 108 in the Ciao dataset.
For user 108, the predicted latent vector by our method is very close to that from MF. As such, the predicted rating by using our method and MF are also very close as shown in Table \ref{table:table7}. For such a user, there is little performance gain to track
concept drift in his/her latent vector.

In Epinions, Flixster and MovieLens datasets, we observe that the latent vectors of most users change smoothly but not usually in a consistent direction. As such, the predicted $P_{i}(T)$ is similar to the original latent factors $P_{i}$. The corresponding latent vectors and the prediction results are shown in Table \ref{table:table8} and the rightmost figure of Figure \ref{fig:2} for a typical user (user 339) in the Epinions dataset. In this case, considering the overall ratings without using the information of time stamps (such as MF) is capable of yielding good estimations and there is little performance gain to
 track concept drift in the user latent vector for most users.
\begin{table}
{\tiny
\begin{center}
\caption{The first five factors of user 49 in the Ciao dataset  and the corresponding prediction results.}
\begin{tabular}{|c||c|c|c|c|c|}
\hline
Factor &  1 &  2 &  3 &  4 &  5 \\ \hline\hline
$P_{i}$ & 0.4592 & 0.2673 & 0.2304 & 0.3172 & 0.3549\\ \hline
$\hat{P}_{i}(1)$ & \multirow{4}{*}{0.4592} & \multirow{4}{*}{0.2673} & \multirow{4}{*}{0.2304} & \multirow{4}{*}{0.3172} & \multirow{4}{*}{0.3549} \\
$\hat{P}_{i}(2)$ & & & & & \\
$\hat{P}_{i}(3)$ & & & & & \\
$\hat{P}_{i}(4)$ & & & & & \\
$\hat{P}_{i}(5)$ & 0.5507 & 0.3000 & 0.2411 & 0.3640 & 0.4328 \\ \hline \hline
 $\hat{P}_{i}(6)$ & 0.5737 & 0.3082 & 0.2438 & 0.3757 & 0.4522\\ \hline
\multicolumn{1}{l}{}\vspace*{0.5cm}
\end{tabular}
\begin{tabular}{|c|c|c|c|c|}
\hline
user ID & item ID & actual rating & MF & Our method \\ \hline\hline
\multirow{5}{*}{49} & 36  & 4 & 3.13  & 3.76 \\ \hhline{~----}
& 138 & 4 & 3.50  & 4.20 \\ \hhline{~----}
& 711 & 5 & 3.11  & 3.74 \\ \hhline{~----}
& 712 & 5 & 3.80  & 4.57 \\ \hhline{~----}
& 713 & 5 & 3.78  & 4.55 \\ \hline
\end{tabular}
%\caption{The user latent vector changes in only one step and the corresponding prediction results for user 49 in the Ciao dataset.}
\label{table:table6}
\end{center}
}
\end{table}

\iffalse
\begin{figure}
\begin{center}
\includegraphics[width=0.8\columnwidth]{./fig/ciao_49}
\caption{The temporal dynamics of the user latent vector for user 49 in the Ciao dataset.}
\label{fig:2}
\end{center}
\end{figure}
\fi

\begin{table}
{\tiny
\begin{center}
\caption{The first five factors of user 108 in the Ciao dataset  and the corresponding prediction results.}
\begin{tabular}{|c||c|c|c|c|c|}
\hline
Factor &  1 &  2 &  3 &  4 &  5 \\ \hline\hline
 $P_{i}$ & 0.5814 & 0.5842 & 0.5738 & 0.5327 & 0.4485\\ \hline
$\hat{P}_{i}(1)$ & 0.5814 & 0.5842 & 0.5738 & 0.5327 & 0.4485 \\ \hline
$\hat{P}_{i}(2)$ & 0.5777 & 0.4566 & 0.4518 & 0.4261 & 0.3749\\ \hline
$\hat{P}_{i}(3)$ & 0.6218 & 0.4630 & 0.4285 & 0.4449 & 0.3949\\ \hline
$\hat{P}_{i}(4)$ & 0.5849 & 0.4845 & 0.4855 & 0.4655 & 0.4043\\ \hline
$\hat{P}_{i}(5)$ & 0.5590 & 0.5398 & 0.4818 & 0.4703 & 0.4174 \\ \hline \hline
$\hat{P}_{i}(6)$ & 0.5534 & 0.5287 & 0.4588 & 0.4547 & 0.4096\\ \hline
\multicolumn{1}{l}{}\vspace*{0.5cm}
\end{tabular}
\begin{tabular}{|c|c|c|c|c|}
\hline
user ID & item ID & actual rating & MF & Our method \\ \hline\hline
\multirow{16}{*}{108} & 122 & 5 & 4.16  & 4.12 \\ \hhline{~----}
& 251 & 4 & 3.57  & 3.59 \\ \hhline{~----}
& 447 & 5 & 4.47  & 4.49 \\ \hhline{~----}
& 469 & 5 & 4.83  & 4.88 \\ \hhline{~----}
& 531 & 5 & 4.49  & 4.57 \\ \hhline{~----}
& 768 & 5 & 5.01  & 5.01 \\ \hhline{~----}
& 823 & 5 & 3.78  & 3.86 \\ \hhline{~----}
& 1258 & 5 & 3.65  & 3.70 \\ \hhline{~----}
& 1319 & 4 & 4.16  & 4.13 \\ \hhline{~----}
& 1320 & 5 & 3.74  & 3.78 \\ \hhline{~----}
& 1321 & 5 & 5.17  & 5.16 \\ \hhline{~----}
& 1322 & 5 & 4.78  & 4.82 \\ \hhline{~----}
& 1323 & 4 & 3.72  & 3.62 \\ \hhline{~----}
& 1324 & 5 & 4.74  & 4.79 \\ \hhline{~----}
& 1325 & 5 & 4.74  & 4.77 \\ \hhline{~----}
& 1326 & 5 & 5.03  & 5.14 \\ \hline
\end{tabular}
\label{table:table7}
\end{center}
}
\end{table}

\iffalse
\begin{figure}
\begin{center}
\includegraphics[width=0.8\columnwidth]{./fig/ciao_108}
\caption{The temporal dynamics of the user latent vector for user 108 in the Ciao dataset.}
\label{fig:3}
\end{center}
\end{figure}
\fi

\begin{table}
{\tiny
\begin{center}
\caption{The first five factors of user 339 in the Epinions dataset  and the corresponding prediction results.}
\begin{tabular}{|c||c|c|c|c|c|}
\hline
Factor &  1 &  2 &  3 &  4 &  5 \\ \hline\hline
$P_{i}$ & 0.4953 & 0.5479 & 0.4734 & 0.5847 & 0.5540 \\ \hline
$\hat{P}_{i}(1)$ & 0.4781 & 0.5194 & 0.4089 & 0.6105 & 0.5400 \\ \hline
$\hat{P}_{i}(2)$ & 0.4810 & 0.5245 & 0.4178 & 0.6249 & 0.5505\\ \hline
$\hat{P}_{i}(3)$ & 0.4968 & 0.5065 & 0.4450 & 0.6246 & 0.5427\\ \hline
$\hat{P}_{i}(4)$ & 0.4793 & 0.5483 & 0.4625 & 0.6249 & 0.5398\\ \hline
$\hat{P}_{i}(5)$ & 0.4825 & 0.5496 & 0.4789 & 0.5966 & 0.5512 \\ \hline \hline
 $\hat{P}_{i}(6)$ & 0.4837 & 0.5571 & 0.4964 & 0.5932 & 0.5540\\ \hline
\multicolumn{1}{l}{}\vspace*{0.5cm}
\end{tabular}
\begin{tabular}{|c|c|c|c|c|}
\hline
user ID & item ID & actual rating & MF  & Our method \\ \hline\hline
\multirow{7}{*}{339} & 9188 & 4 & 3.57  & 3.59 \\ \hhline{~----}
& 9189 & 1 & 0.95  & 0.96 \\ \hhline{~----}
& 9190 & 4 & 3.38  & 3.39 \\ \hhline{~----}
& 9191 & 4 & 3.93  & 3.93 \\ \hhline{~----}
& 9192 & 4 & 3.54  & 3.56 \\ \hhline{~----}
& 9193 & 5 & 3.98  & 4.11 \\ \hhline{~----}
& 9194 & 3 & 4.02  & 4.03 \\ \hline
\end{tabular}
%\caption{The evolution of user preferences and corresponding prediction results in Epinions dataset.}
\label{table:table8}
\end{center}
}
\end{table}

\iffalse
\begin{figure}
\begin{center}
\includegraphics[width=0.8\columnwidth]{./fig/epinions_339}
\caption{The temporal dynamics of the user latent vector for user 339 in the Epinions dataset.}
\label{fig:4}
\end{center}
\end{figure}
\fi

At the end of this section, we report the run-time  of our temporal method on these datasets in Table \ref{table:table9}.
% and Figure \ref{fig:5}.
The run-time includes learning the user latent vectors, learning the transition matrices, and further performing rating prediction which quantify the additional efforts after obtaining the original latent matrices $P$ and $Q$ from MF. As we use Matlab to implement our temporal method (except we use LIBMF \cite{chin2015fast, chin2015learning} in the step for MF), we also
 implement MF by using Matlab and report the run-time for performing MF on these datasets. As shown in Table \ref{table:table9}, the run-time of MF by LIBMF is in the order of seconds and the run-time of TMF and MF by Matlab  is in the order of minutes. The additional efforts of our temporal methods (in terms of run-time) are comparable to those for performing MF by using Matlab.

\iffalse
In addition to the amount of training data, the number of iterations needed for the SGD algorithm to converge is another key factor to the running time. The number of iterations needed in the Ciao dataset are not only more than those in the other three real datasets but also larger than the settings in original MF. Based on this observation and the overall RMSE in Table \ref{table:table5}, it is possible that the number of iterations of the stochastic gradient descent algorithm could be a useful indicator to reveal whether the temporal effect is significant in a dataset.
\fi
%\iffalse
\begin{table}
{\tiny
\begin{center}
\caption{The run-time for our temporal method and MF on various datasets.}
\begin{tabular}{|c|c|c|c|c|c|}
\hline
& Synthetic & Ciao & Epinions & Flixster & MovieLens \\ \hline\hline
TMF  & 57.60m & 3.79m & 25.90m & 78.42m & 143.08m \\ \hline
MF (LIBMF)  & 2.32s & 0.18s & 6.85s & 20.47s & 44.27s \\ \hline
MF (Matlab)  & 27.61m & 0.67m & 24.58m & 218.75m & 532.93m \\ \hline
%Iterations & 30 & 68 & 13 & 1 & 1\\ \hline
\end{tabular}
\label{table:table9}
\end{center}
}
\end{table}

\section{Conclusions}
\label{c:conclusions}

%The ability of tracking concept drift in each individual user latent vector might play a crucial role for recommender systems to  better serve %``valuable'' customers in the future.
%Motivated by this, 
In this paper, we proposed  a Temporal Matrix Factorization approach (TMF) for tracking concept drift in each individual user latent vector. There are two key innovative steps in our approach: (i) a modified stochastic gradient descent method to learn an individual user latent vector at each time step, and (ii) a linear model for the transition of the individual user latent vectors by the Lasso regression.
In comparison with the other approaches that intend to incorporate temporal dynamics with MF in the literature, there are several distinctive features of our temporal method:
%\begin{description}

\noindent (i) In comparison with timeSVD++ \cite{koren2010collaborative}, our systematic approach is more structured and does not require fine tuning a lot of unstructured parameters.

\noindent (ii) Our modified stochastic gradient descent method is able to alleviate the data sparsity problem for learning the user preferences at a certain time step. This overcomes the data sparsity problem in tensor factorization.

\noindent (iii) Unlike the CKF approach \cite{sun2014collaborative}, we do not need to assume the transition matrix is {\em homogeneous}. Thus, we are allowed to track concept drift in each individual user latent vector. 
%\end{description}

        In comparison with the original MF,  our temporal method is able to achieve lower root mean square errors (RMSE) for both the synthetic and real datasets. One interesting finding is that the performance gain in RMSE is mostly from those users
who indeed have concept drift in their user latent vectors at the time of prediction.
As our temporal method is specifically designed for each user,  one can save a lot of efforts by only tracking those users who indeed have concept drift in their user latent vectors at the time of prediction. However, identifying
those users is not an easy task and might require further study. One possible approach for this is to examine the transition matrix for each user. In our experiments, we found that there are many users whose transition matrices are the identity matrix and those users are not worth tracking.
 
%Not every customer needs to be tracked. 

Another research direction is to study the effect of cold start users (who have very few ratings). One might think cold start users are difficult to predict and then immediately filter out their ratings in the preprocessing step. However, in our temporal method,
the ratings of cold start users might be valuable as they contribute to the item latent matrix $Q$ which in turn affects the accuracy of estimating the time series of the latent vectors of other  users.

\iffalse
In this paper, we study the rating prediction problem which is the core in recommender systems. We proposed a principled and general MF-based approach to incorporate the temporal dynamics into recommendation. By adequately utilizing stochastic gradient descent method to capture the evolution of user preferences and learning the transition patterns at individual level with Lasso regression, our temporal dynamic model is able to achieve more accurate predictions compared to the MF approach by 20\% on the synthetic dataset and 5\% on Ciao dataset. Besides, we discuss the characteristics of evolution which are the main reason for extent of advancement on the different real datasets. In future work, we can incorporate additional information such as social network analysis or content-based features into the temporal dynamic model based on the well-structured MF-based framework we developed. Moreover, this comprehensive model with a variety of information is more capable of addressing the cold start problem which is another aspect of future research direction.
\fi

%\bibliographystyle{unsrt}
\bibliographystyle{abbrv}
%\bibliography{thesisbib}

\end{document}